\documentclass[aps,pra,twocolumn,superscriptaddress]{revtex4-1}

\usepackage{graphicx,subfigure}
\usepackage{amsmath,amssymb,bm,amsthm}
\usepackage{mathtools}
\usepackage{multirow}
\usepackage{textcomp}
\usepackage{float}
\usepackage{xcolor}
\usepackage[normalem]{ulem}
\usepackage{dsfont}
\usepackage{url}

\newcommand{\ave}[1]{\left \langle #1 \right \rangle}
\newcommand{\ket}[1]{\left | #1 \right \rangle}
\newcommand{\bra}[1]{\left \langle #1 \right |}
\newcommand{\mele}[3]{\left \langle #1 \middle | #2 \middle | #3 \right \rangle}

\newcommand{\tr}{\text{Tr}}
\newcommand{\oph}{\hat{H}}
\newcommand{\opn}{\hat{n}}
\newcommand{\opphi}{\hat{\phi}}
\newcommand{\opu}{\hat{U}}
\newcommand{\ddt}{\frac{\mathrm{d}}{\mathrm{d} t}}

\begin{document}

\title{Quantum control and noise protection of a Floquet $0-\pi$ qubit}

\author{Zhaoyou Wang}
\email{zhaoyou@stanford.edu}
\altaffiliation[Current address: ]{Pritzker School of Molecular Engineering, The University of Chicago, Chicago, Illinois 60637, USA}
\affiliation{E. L. Ginzton Laboratory and the Department of Applied Physics, Stanford University, Stanford, CA 94305 USA}

\author{Amir H. Safavi-Naeini}
\email{safavi@stanford.edu}
\affiliation{E. L. Ginzton Laboratory and the Department of Applied Physics, Stanford University, Stanford, CA 94305 USA}
\affiliation{AWS Center for Quantum Computing, Pasadena, CA, 91125, USA}

\date{\today}

\begin{abstract}
Time-periodic systems allow engineering new effective Hamiltonians from limited physical interactions. For example, the inverted position of the Kapitza pendulum emerges as a stable equilibrium with rapid drive of its pivot point.
In this work, we propose the \emph{Kapitzonium}: a Floquet qubit that is the superconducting circuit analog of a mechanical Kapitza pendulum.
Under periodic driving, the emerging qubit states are exponentially protected against bit and phase flips caused by dissipation, which is the primary source of decoherence of current qubits. However, we find that dissipation causes leakage out of the Floquet qubit subspace.
We engineer a passive cooling scheme to stabilize the qubit subspace, which is crucial for high fidelity quantum control under dissipation. 
Furthermore, we introduce a hardware-efficient fluorescence-based method for qubit measurement and discuss the experimental implementation of the Floquet qubit.
The proposed Kapitzonium is one of the simplest Floquet qubits that can be realized with current technology -- and it already has many intriguing features and capabilities. Our work provides the first steps to develop more complex Floquet quantum systems from the ground up to realize large-scale protected engineered dynamics.
\end{abstract}

\maketitle

\section{Introduction}
Superconducting circuits offer flexible qubit designs, which hold promise for engineering scalable quantum computers~\cite{blais2021,kjaergaard2020,koch2007,manucharyan2009,brooks2013,gyenis2021a,kalashnikov2020,pechenezhskiy2020,ofek2016,hu2019,campagne-ibarcq2020}.
Quantum information is encoded in eigenstates of the circuit, and qubit properties can be engineered with different circuit designs.
Most superconducting qubits today are based on single-node circuits, which have demonstrated long coherence time and high gate fidelity~\cite{blais2021,kjaergaard2020,koch2007,manucharyan2009}.
More complex circuits have been proposed, such as the $0-\pi$ qubit~\cite{brooks2013}, which provide a greater level of protection to decay and dephasing. The intrinsic noise protection of these circuits comes from engineering two degenerate ground states with disjoint wavefunctions~\cite{groszkowski2018,gyenis2021}. Many local noise processes are significantly suppressed by these types of states.
However, the protected qubits usually require circuit parameters that are demanding for current experiments~\cite{groszkowski2018,gyenis2021a,gyenis2021}.

Alternative and less demanding approaches to realizing circuit-level noise protection has emerged over the last decade. In these approaches, time-modulation of the superconducting circuit is used to engineer a subspace that is protected against either bit/phase flips or both. Prominent examples of these schemes include the dissipative cats~\cite{mirrahimi2014,leghtas2015,lescanne2020}, Kerr cats~\cite{puri2017,grimm2020}, as well as recent proposals and implementations of autonomously corrected qubits~\cite{kapit2016,li2023}.  In all of these works, the modulation frequencies and amplitudes are tuned to induce a certain set of transitions. The drive amplitudes must also be limited to avoid driving higher-order or  unwanted transitions.  

Here we propose a qubit design based on circuits that are periodically modulated in time. The modulation in our case is highly nonperturbative and not tuned to any particular frequency.
In general, the dynamics of a time-periodic system is governed by the effective Hamiltonian from the Floquet theory~\cite{eckardt2017,venkatraman2022}. Engineering the Floquet Hamiltonian enables new designs of superconducting qubits~\cite{didier2019,huang2021,gandon2022}.
The Floquet qubit we study is the superconducting circuit analog to the mechanical Kapitza pendulum~\cite{landau1976}. In Kapitza pendulum, the pivot point of the pendulum is periodically moved up and down (Fig.~\ref{fig:fig1}(a)), leading to qualitatively new dynamics. Notably,  Kapitza pendulum has two stable equilibria: one at $\phi=0$ and the other at $\phi=\pi$. In the qubit that we propose here, these correspond to the qubit basis states.
We thus name this Floquet $0-\pi$ qubit the \emph{Kapitzonium}.

Our paper is organized as follows. In Sec.~\ref{main_section_2}, we introduce Kapitzonium and its unitary gates.
In Sec.~\ref{main_section_3}, we consider the open system dynamics of Kapitzonium where heating effects induced by charge noise are suppressed with engineered cooling.
In Sec.~\ref{main_section_4}, we discuss some technical aspects towards the experimental implementation of the Kapitzonium. In Sec.~\ref{main_section_5}, we conclude our discussion with potential directions for future study.

\section{Unitary dynamics of Kapitzonium}
\label{main_section_2}

\subsection{Kapitzonium Hamiltonian}
\begin{figure}[t]
    \centering
    \includegraphics[width=0.48\textwidth]{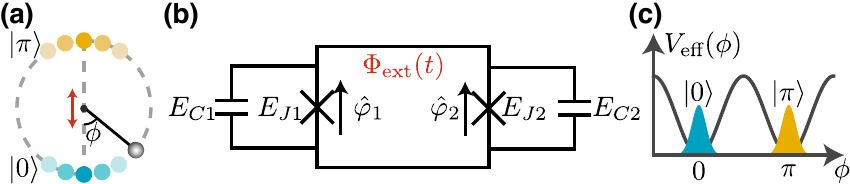}
    \caption{Schematics of Kapitzonium. (a) The Kapitza pendulum. (b) Superconducting circuit implementing the Kapitzonium. (c) Effective double well potential showing the disjoint wavefunctions and degeneracy of $\ket{0}$ and $\ket{\pi}$.}
    \label{fig:fig1}
\end{figure}

The Kapitzonium circuit (Fig.~\ref{fig:fig1}(b)) is identical to that of a  capacitively shunted superconducting quantum interference device (SQUID). The external flux $\Phi_{\text{ext}}(t)$ threading the SQUID loop is used to emulate the effect of a time-modulated pivot point of a pendulum.
The Kapitzonium Hamiltonian is derived with the established circuit quantization procedures~\cite{vool2017,you2019,rajabzadeh2022,chitta2022}.
The branch flux variable $\hat{\varphi}_k$ across each Josephson junction and the conjugate charge variable $\opn_k$ satisfy $[\hat{\varphi}_k,\opn_k] = i$ for $k=1,2$. The circuit Hamiltonian is
\begin{equation}
    \label{eq:full_circuit_Hamiltonian}
    \oph = \sum_{k=1,2} 4 E_{Ck} \opn^2_k - E_{Jk} \cos \hat{\varphi}_k ,
\end{equation}
where $E_{Jk}$ are the Josephson energies and $E_{Ck}=e^2/2 C_k$ are the charging energies with $C_k$ being the junction capacitances.

Flux quantization sets the constraint of $\hat{\varphi}_1 - \hat{\varphi}_2 = \Phi_{\text{ext}}(t)$.
In the symmetric case of $E_{J1}=E_{J2}$ and $E_{C1}=E_{C2}$, Eq.~(\ref{eq:full_circuit_Hamiltonian}) reduces to the Kapitzonium Hamiltonian
\begin{equation}
    \label{eq:Hamiltonian_general}
    \oph(t) = 4 E_C \opn^2 - E_J \cos \phi_{\text{ext}} (t) \cos \opphi
\end{equation}
where $\phi_{\text{ext}}(t) = \Phi_{\text{ext}}(t)/2$. The flux variable $\hat{\phi} = (\hat{\varphi}_1+\hat{\varphi}_2)/2$ corresponds to the pendulum rotation angle, and $\opn=\opn_1 + \opn_2$ is the conjugate charge variable satisfying $[\opphi,\opn]=i$. The charging energy is $E_C=e^2/2(C_1+C_2)$ and the Josephson energy is $E_J = E_{J1}+E_{J2}$.

To realize the analog of the Kapitza pendulum, we set the external flux to $\phi_{\text{ext}}(t)=\omega t$.
The idling dynamics are described by a Floquet Hamiltonian
\begin{equation}
    \label{eq:Hamiltonian_idle}
    \oph_0(t) = 4 E_C \opn^2 - E_J \cos \omega t \cos \opphi
\end{equation}
with a time period of $T=2\pi/\omega$.
Throughout this paper, we choose Kapitzonium parameters $E_J/2\pi=100~\text{GHz}$, $\omega/2\pi=10~\text{GHz}$ and $E_C/2\pi=0.01~\text{GHz}$ unless otherwise specified. To understand the emergence of the degenerate ``ground'' states, we derive the effective Hamiltonian~\cite{eckardt2017,venkatraman2022} of $\oph_0(t)$ by integrating over the fast modulation, and find
\begin{equation}
    \label{eq:effective_Hamiltonian}
    \hat{H}_{\text{eff}} = 4E_C \hat{n}^2 - \tilde{E}_J \cos 2\hat{\phi} ,
\end{equation}
where $\tilde{E}_J=E_C E_J^2/\omega^2$.
Here we only keep terms up to the second order in the effective Hamiltonian, while the third order term is 0 and the fourth order term scales as $E_C^3 E_J^2 / \omega^4 \ll \tilde{E}_J$, which can be neglected~\cite{eckardt2017,venkatraman2022}.

The static effective Hamiltonian $\hat H_{\text{eff}}$ possesses some of the key features of a protected qubit: disjoint wavefunctions and energy degeneracy.
Since $\opphi$ is $2\pi$-periodic, $\hat{H}_{\text{eff}}$ has two near-degenerate ground states $(\ket{0} \pm \ket{\pi} ) / \sqrt{2}$ in the deep transmon regime of $\tilde{E}_J / E_C = 100$. Here $\ket{0}$ and $\ket{\pi}$ are localized at the two minima of the effective potential $V_{\text{eff}} (\phi) =  - \tilde{E}_J \cos 2\phi$ (Fig.~\ref{fig:fig1}(c)) with exponentially small overlap.

\begin{figure}[t]
    \centering
    \includegraphics[width=0.48\textwidth]{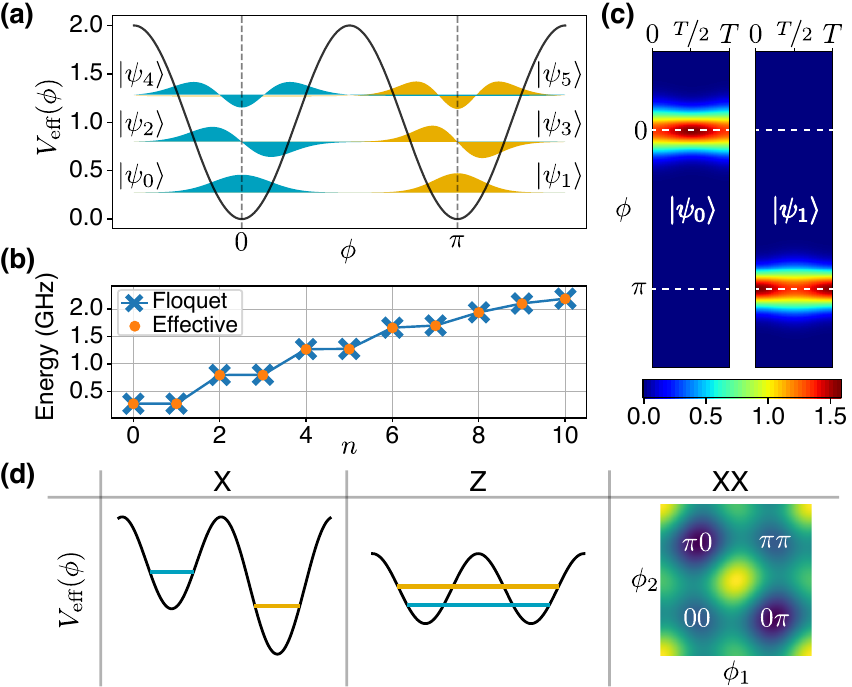}
    \caption{(a) Floquet eigenstates of $\oph_0(t)$. (b) Floquet eigenenergies $\oph_0(t)$ compared with the spectrum of $\oph_{\text{eff}}$. (c) Time evolution of $\ket{\psi_0}$ and $\ket{\psi_1}$ from $0$ to $T$. Here we plot the probability distribution of the $\phi$ space wavefunction. (d) The effective potential when performing different gates.}
    \label{fig:fig2}
\end{figure}

\subsection{Floquet eigenstates}
The effective Hamiltonian provides a useful approximate picture of the potential and the ``ground'' states of the Kapitzonium -- note that the states $(\ket{0} \pm \ket{\pi} ) / \sqrt{2}$ are only the ground states of the effective Hamiltonian. To better understand the states, gates, and noise, we will need to move beyond the effective description and consider the Floquet eigenstates of the qubit.
The Floquet eigenstates $\ket{\Psi_\alpha (t)}$ of the idling Kapitzonium Hamiltonian $\oph_0(t)$ satisfy
\begin{equation}
    \oph_0(t) \ket{\Psi_\alpha (t)} = i \frac{d}{dt} \ket{\Psi_\alpha (t)} ,
\end{equation}
and have the form of
\begin{equation}
    \ket{\Psi_\alpha (t)} = e^{-i \varepsilon_\alpha t} \ket{\Phi_\alpha (t)} .
\end{equation}
Here $\ket{\Phi_\alpha (t)} = \ket{\Phi_\alpha (t+T)}$ are the periodic Floquet modes and $\varepsilon_\alpha$ are the Floquet eigenenergies.

In Fig.~\ref{fig:fig2}(a), we plot the even and odd superpositions of the Floquet eigenstates defined as
\begin{equation}
    \begin{split}
        \ket{\psi_{2k}} =& \frac{1}{\sqrt{2}} \left( \ket{\Psi_{2k}(t=0)}+\ket{\Psi_{2k+1}(t=0)} \right) \\
        \ket{\psi_{2k+1}} =&
        \frac{1}{\sqrt{2}} \left( \ket{\Psi_{2k}(t=0)}-\ket{\Psi_{2k+1}(t=0)} \right)
    \end{split}
\end{equation}
for $k=0,1,2$. The phases of $\ket{\Psi_\alpha(t=0)}$ are chosen such that $\ket{\psi_{2k}}$ ($\ket{\psi_{2k+1}}$) are mostly localized around $\phi=0$ ($\phi=\pi$).
In Fig.~\ref{fig:fig2}(b), we compare the exact Floquet eigenenergies $\varepsilon_\alpha$ with the spectrum of $\oph_{\text{eff}}$ and find a close match.
Here the index $\alpha$ are sorted based on the overlaps between $\ket{\Psi_\alpha(t=0)}$ and the $n$th eigenstates of $\oph_{\text{eff}}$.
The disjoint wavefunctions of $\{ \ket{\psi_{2k}}, \ket{\psi_{2k+1}} \}$ and the near degeneracy $\varepsilon_{2k} \approx \varepsilon_{2k+1}$ for $k=0,1,2$ confirm the effective double well potential $V_{\text{eff}} (\phi)$.
Finally, Floquet eigenstates are not stationary within one period.
For example, the time evolution of $\ket{\psi_0}$ and $\ket{\psi_1}$ from $0$ to $T$ shows varying $\phi$ space wavefunctions (Fig.~\ref{fig:fig2}(c)).

A single qubit state $(c_0,c_1)^T$ can be encoded in Kapitzonium as
\begin{equation}
    \ket{\psi(t)} = c_0 \ket{\Psi_0 (t)} + c_1 \ket{\Psi_1 (t)} ,
\end{equation}
where $c_0$ and $c_1$ are invariant under $\oph_0(t)$.
For $(\varepsilon_1-\varepsilon_0) /2\pi \approx 4.7$~kHz, $1/(\varepsilon_1-\varepsilon_0)$ is much longer than the time scale of any relevant Kapitzonium operations. We thus assume $\varepsilon_0 = \varepsilon_1$ and only consider the system at $t=nT$ to simplify our discussions.
In this case, the qubit basis states $\ket{\Psi_0}$ and $\ket{\Psi_1}$ become static, and are related to $\ket{0}$ and $\ket{\pi}$ by a Hadamard transformation
\begin{equation}
    \begin{split}
        \ket{\Psi_0} =& \frac{1}{\sqrt{2}} \left( \ket{0} + \ket{\pi} \right) \\
        \ket{\Psi_1} =& \frac{1}{\sqrt{2}} \left( \ket{0} - \ket{\pi} \right) .
    \end{split}
\end{equation}

\subsection{Kapitzonium gates}
The encoded qubit state can be manipulated by engineering the flux drive $\phi_{\text{ext}} (t)$. The resulting effective potentials (Fig.~\ref{fig:fig2}(d)) provide the intuition for the Kapitzonium gates. Here we focus on the gate Hamiltonians and details on the required flux drive are discussed in Sec.~\ref{main_section_4}.

\textbf{X rotation}.
The X gate Hamiltonian $\oph_x(t) = \oph_0(t) + \alpha_x \cos \opphi$ generates the rotation along X axis.
The effective potential now becomes an asymmetric double well $V_{\text{eff}}^{(x)} (\phi) =  - \tilde{E}_J \cos 2\phi + \alpha_x \cos \phi$, which lifts the degeneracy between $\ket{0}$ and $\ket{\pi}$.
Therefore, in the $ \{ \ket{0},\ket{\pi} \}$ basis $\oph_x(t) \approx \alpha_x \hat{\sigma}_z$ and in the $\{ \ket{\Psi_0}, \ket{\Psi_1} \}$ basis, $\oph_x(t) \approx \alpha_x \hat{\sigma}_x$, leading to Rabi oscillation between $\ket{\Psi_0}$ and $\ket{\Psi_1}$.

\textbf{Z rotation}.
The depth $\tilde{E}_J$ of the effective double well potential can be controlled dynamically with the flux driving frequency $\omega$. Increasing $\omega$ reduces $\tilde{E}_J / E_C = (E_J/\omega)^2$ and induces stronger coupling between $\ket{0}$ and $\ket{\pi}$.
We choose $\omega_z/2\pi=20$~GHz for the Z gate, which lifts the degeneracy between $\ket{\Psi_0}$ and $\ket{\Psi_1}$ with a splitting $(\varepsilon_1 - \varepsilon_0)/2\pi \approx 1.8$~MHz.
This implements a phase gate for $\ket{\Psi_0}$ and $\ket{\Psi_1}$, i.e., the rotation along Z axis.

\textbf{Two qubit XX rotation}.
We couple two Kapitzonium with a Josephson junction to realize the XX gate Hamiltonian $\oph_{xx}(t) = \oph_0(t) + \alpha_{xx} \cos (\hat{\phi}_1 - \hat{\phi}_2)$. Replacing the Josephson junction with a SQUID makes the coupling tunable.
The joint effective potential is $V_{\text{eff}}^{(xx)} (\phi_1,\phi_2) =  - \tilde{E}_J \cos 2\phi_1 - \tilde{E}_J \cos 2\phi_2 + \alpha_{xx} \cos (\phi_1 - \phi_2)$, which lifts the degeneracy between $\{ \ket{00},\ket{\pi\pi} \}$ and $\{ \ket{0\pi},\ket{\pi 0} \}$.
In the $ \{ \ket{0},\ket{\pi} \}$ basis $\oph_{xx}(t) \approx \alpha_{xx} \hat{\sigma}_z \otimes \hat{\sigma}_z$, and in the $\{ \ket{\Psi_0}, \ket{\Psi_1} \}$ basis $\oph_{xx}(t) \approx \alpha_{xx} \hat{\sigma}_x \otimes \hat{\sigma}_x$, generating the XX rotation.

\section{Open system dynamics of Kapitzonium}
\label{main_section_3}

\subsection{Heating problem}
In an open quantum system, a coupling between the system and bath allows the bath to induce transitions between different eigenstates of the system. The form of the coupling and the temperature of the bath both go into determining the transitions and their rates. For simplicity, we only consider a zero-temperature bath in this paper. For time-independent systems, any transition from lower to higher energy will require energy to be absorbed from the bath. Because of this, static systems coupled to  zero-temperature baths eventually decay to their ground states.

Considering the level structure in Fig.~\ref{fig:fig2}(a), we would naively expect the zero-temperature bath to induce only the downward transition $\ket{\Psi_2} \rightarrow \ket{\Psi_0}$, with the opposing transition $\ket{\Psi_0} \rightarrow \ket{\Psi_2}$ being suppressed as that would require absorption of energy from the bath. This intuition is incorrect because the Kapitzonium is a Floquet system, which is constantly exchanging energy with the Floquet drive. As a result, the bath induced transitions happen in both directions between the Floquet eigenstates: both $\ket{\Psi_2} \rightarrow \ket{\Psi_0}$ and $\ket{\Psi_0} \rightarrow \ket{\Psi_2}$ transitions are allowed in a Kapitzonium interacting with a zero-temperature bath.
Therefore in contrast to the static $0-\pi$ qubit~\cite{brooks2013}, the Floquet $\ket{0}$ and $\ket{\pi}$ can still \emph{decay} out of the qubit subspace through what looks like a heating process.

Consider a generic Floquet system $\oph_0(t) = \oph_0 (t+T)$ coupled to some bath degrees of freedom $\hat{B}(t)$ via the system operator $\hat{O}$.
The system-bath Hamiltonian in the rotating frame of the bath is
\begin{equation}
    \oph_{\text{SB}}(t) = \oph_0 (t) + \hat{O} \hat{B}(t) ,
\end{equation}
where
\begin{equation}
    \hat{B}(t) = \sum_k g_k \left( \hat{b}_k e^{-i\omega_k t} + \hat{b}_k^\dagger e^{i\omega_k t} \right) .
\end{equation}
Here $\omega_k \geq 0$ is the frequency of the $k$th bath mode and $g_k$ is the coupling between the $k$th mode and the system.

The emission spectrum of the Floquet system can be calculated in the interaction picture of $\oph_0(t)$, which we call the Floquet frame.
More specifically, the unitary $\opu_0(t)$ generated by $\oph_0(t)$ is given by
\begin{equation}
    \opu_0(t,t_0) = \sum_{\alpha} \ket{\Psi_\alpha (t)} \bra{\Psi_\alpha (t_0)} ,
\end{equation}
where $\ket{\Psi_\alpha (t)}$ are the Floquet eigenstates of $\oph_0(t)$, and $\opu_0(t,t_0)$ satisfies the Schr\"{o}dinger equation
\begin{equation}
     i \frac{d}{d t} \opu_0 (t,t_0) = \hat{H}_0(t) \opu_0(t,t_0).
\end{equation}
Now we could perform the unitary transformation $\opu_0 (t,t_0)$ to enter the Floquet frame, where the system-bath Hamiltonian becomes
\begin{equation}
    \tilde{H}_{\text{SB}}(t) = \hat{O} (t) \hat{B}(t) ,
\end{equation}
and
\begin{equation}\label{eq:op_expansion}
    \hat{O}(t) = \opu_0^\dagger (t,t_0) \hat{O} \opu_0(t,t_0) = \sum_{\alpha \beta} O_{\alpha \beta}(t) \ket{\Psi_\alpha (t_0)} \bra{\Psi_\beta (t_0)}
\end{equation}
with $O_{\alpha \beta}(t) = \mele{\Psi_\alpha (t)}{\hat{O}}{\Psi_\beta (t)}$. We set $t_0=0$ without loss of generality.

Since the Floquet modes $\ket{\Phi_\alpha (t)}$ are periodic in time, we Fourier expand $O_{\alpha \beta}(t)$:
\begin{equation}
    \begin{split}
        O_{\alpha \beta}(t) =& e^{i (\varepsilon_\alpha - \varepsilon_\beta) t} \mele{\Phi_\alpha(t)}{\hat{O}}{\Phi_\beta(t)} \\
        =& e^{i (\varepsilon_\alpha - \varepsilon_\beta) t} \sum_{n=-\infty}^{\infty} O_{\alpha\beta n} e^{i n \omega t} ,
    \end{split}
\end{equation}
where $O_{\alpha\beta n}$ together with $\{\varepsilon_\alpha\}$ gives the emission spectrum.
Since all $\hat{b}_k$ modes are in vacuum, transition $\alpha \rightarrow \beta$ from $\ket{\Psi_\alpha (t_0)}$ to $\ket{\Psi_\beta (t_0)}$ is only possible if $\varepsilon_\alpha - \varepsilon_\beta + n \omega > 0$, with the transition rate determined by $|O_{\alpha\beta n}|^2$ and the bath spectral density at frequency $\varepsilon_\alpha - \varepsilon_\beta + n \omega$.
Furthermore, transition $\alpha \rightarrow \beta$ could emit photons at multiple frequencies, and both $\alpha \rightarrow \beta$ and $\beta \rightarrow \alpha$ could occur, which is different from the relaxation of static systems.

\begin{figure}[t]
    \centering
    \includegraphics[width=0.48\textwidth]{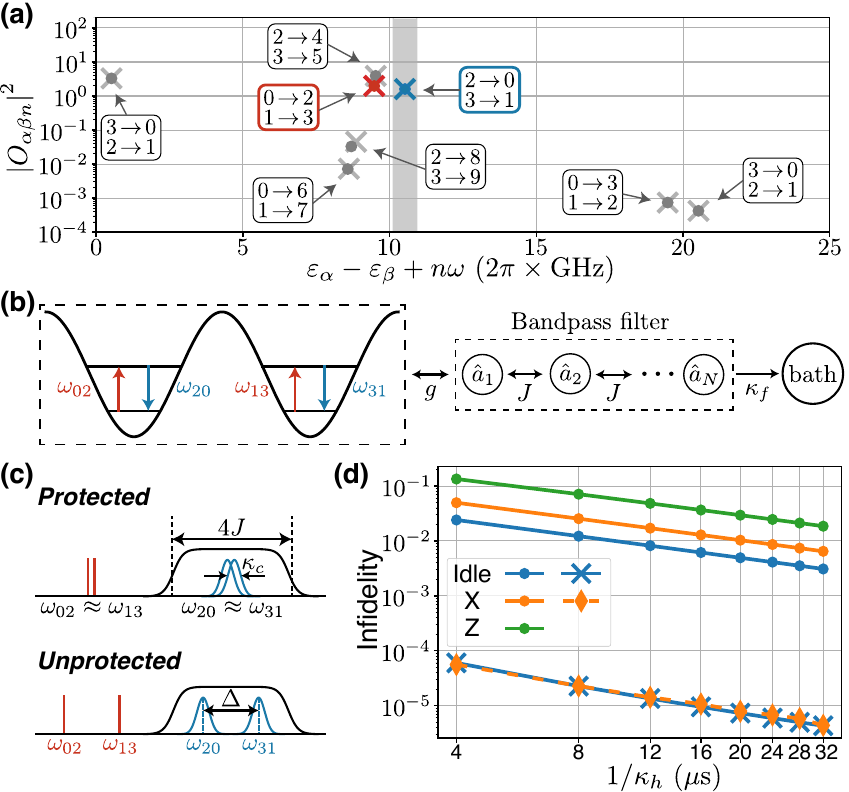}
    \caption{(a) Emission spectrum of Kapitzonium under charge noise. The upper (lower) $m \rightarrow n$ label within each rounded rectangle corresponds to the cross (dot). (b) Schematic for Kapitzonium coupled to a bandpass filter. (c) Two different regimes of the emission spectrum, which determines whether the qubit can be protected or not. (d) Idling and gate fidelity for different $T_1$ time $1/\kappa_h$, with (cross and diamond) and without (dot) the filter.}
    \label{fig:fig3}
\end{figure}

In Fig.~\ref{fig:fig3}(a), we plot $|O_{\alpha\beta n}|^2$ for various transitions of the Kapitzonium under charge noise with $\hat{O}=\hat{n}$.
The dominant heating processes $0 \rightarrow 2$ (red cross) and $1 \rightarrow 3$ (red dot) occur at near degenerate frequency around $\omega_{02}/2\pi \approx \omega_{13} /2\pi \approx 9.5~\text{GHz}$.
The reverse cooling processes $2 \rightarrow 0$ (blue cross) and $3 \rightarrow 1$ (blue dot) are also allowed at a different frequency around $\omega_{20}/2\pi \approx \omega_{31}/2\pi \approx 10.5~\text{GHz}$.

Assuming the bath spectral density is flat, we trace out the bath degree of freedoms and derive a master equation for the Kapitzonium.
In the Floquet frame, the master equation is $\dot{\hat{\rho}} = \kappa_h D[\hat{O}_-(t)] (\hat{\rho})$ (see Appendix~\ref{SI_master_eq}),  where $D[\hat{A}] (\hat{\rho}) = \hat{A} \hat{\rho} \hat{A}^\dagger - \frac{1}{2} \{ \hat{A}^\dagger \hat{A}, \hat{\rho} \}$.
Here $\kappa_h$ is the heating rate due to intrinsic loss and 
\begin{equation}\label{eq:neg_frequency}
    \hat{O}_-(t) = \sum_{\alpha \beta} O_{\alpha \beta}^- (t) \ket{\Psi_\alpha (t_0)} \bra{\Psi_\beta (t_0)} ,
\end{equation}
where
\begin{equation}
    O_{\alpha \beta}^-(t) = e^{i (\varepsilon_\alpha - \varepsilon_\beta) t} \sum_{\varepsilon_\alpha - \varepsilon_\beta +n \omega < 0} O_{\alpha\beta n} e^{i n \omega t} .
\end{equation}
Without the Floquet driving, $\kappa_h$ is approximately the amplitude damping rate of a transmon corresponding to its $1/T_1$.

\subsection{Enhanced cooling with filter}
The frequency dependence of the cavity emission suggests that we can enhance a specific transition rate by increasing the bath spectral density at the transition frequency~\cite{murch2012,putterman2022}.
More concretely, we propose to capacitively couple the Kapitzonium to a bandpass filter around 10.5~GHz with 800~MHz bandwidth (Fig.~\ref{fig:fig3}(a) grey shade region).
The bandpass filter enhances the cooling processes without causing extra heating, which preserves the qubit basis states $\{ \ket{\Psi_0}, \ket{\Psi_1} \}$.
Furthermore, the environment cannot distinguish whether the emitted photon comes from the $2 \rightarrow 0$ or $3 \rightarrow 1$ transition since $\omega_{20} \approx \omega_{31}$.
Therefore the phase coherence between $\ket{\Psi_0}$ and $\ket{\Psi_1}$ is also preserved by the cooling processes, enabling fully autonomous protection of the qubit subspace.

The bandpass filter can be modeled as a chain of linearly coupled harmonic ocsillators $\hat{a}_1 ,..., \hat{a}_N$~\cite{putterman2022}.
The first filter mode $\hat{a}_1$ also couples capacitively to the Kapitzonium via the interaction $g \opn \left( \hat{a}_1 + \hat{a}_1^\dagger \right)$.
The full Hamiltonian is (Fig.~\ref{fig:fig3}(b))
\begin{equation}\label{eq:H_with_filter}
    \begin{split}
        \oph(t) =& \oph_0(t)+ \omega_f \sum_{k=1}^{N} \hat{a}_k^\dagger \hat{a}_k + g \opn \left( \hat{a}_1 + \hat{a}_1^\dagger \right) \\
        & + J \sum_{k=1}^{N-1} \left( \hat{a}_k \hat{a}_{k+1}^\dagger + \hat{a}_k^\dagger \hat{a}_{k+1} \right) ,
    \end{split}
\end{equation}
where $\omega_f$ is the center frequency of the filter and $J$ is the coupling rate between two adjacent filter modes.
The last filter mode $\hat{a}_N$ decays into a zero temperature bath at rate $\kappa_f$, which is described by the Lindblad dissipator $\kappa_f D[\hat{a}_N]$.

The qubit subspace is autonomously protected when the engineered cooling rate is sufficiently larger than the intrinsic heating rate $\kappa_h$ of Kapitzonium.
Following Ref.~\cite{putterman2022}, we choose $N=3$, $\kappa_f=2J$ and $g=\kappa_f/5$, such that the filter bandwidth is $2\kappa_f$ and the filter modes are only weakly excited.
The cooling rate is about $\kappa_c = 4g^2/\kappa_f=4\kappa_f/25$ after adiabatically eliminating all filter modes~\cite{reiter2012,chamberland2022}. For $\kappa_f/2\pi=400$~MHz we have $\kappa_c/2\pi=64$~MHz.
In Fig.~\ref{fig:fig3}(d), we plot the average fidelity of Kapitzonium idling for 50~ns with (blue cross) and without (blue dot) the filter for different values of $1/\kappa_h$.
The cooling enhanced by the filter has reduced the idling infidelity by more than 2 orders of magnitude.

\subsection{Gate protection}
The filter that protects idling may not protect the Kapitzonium gates since the emission spectrum could be different during the gates.
The filter performance is determined by two quantities of the emission spectrum.
One quantity is the frequency spacing $\mathcal{B} \equiv |(\omega_{20}+\omega_{31})-(\omega_{02}+\omega_{13})|/2$ between the dominant heating and cooling transitions. $\mathcal{B}$ limits the filter bandwidth $\kappa_f$ and thus the maximal cooling rate $\kappa_c$.
The other quantity is the degeneracy of the two heating (cooling) transitions measured by $\Delta \equiv |\omega_{02} - \omega_{13}|=|\omega_{20} - \omega_{31}|$.
In the degenerate regime where $\Delta \ll \kappa_c$, such as idling with $\Delta/2\pi \approx 0.2$~MHz, the qubit subspace is protected (Fig.~\ref{fig:fig3}(c)).
In the non-degenerate regime where $\Delta \gtrsim \kappa_c$, the environment could distinguish which cooling transition the emitted photon comes from and dephase the qubit (Fig.~\ref{fig:fig3}(c)).
Therefore the qubit basis states $\ket{\Psi_0}$ and $\ket{\Psi_1}$ are protected but not their coherent superpositions.

For X gate with relatively small $\alpha_x/2\pi \approx 5.2$~MHz, the emission spectrum is similar to the idling emission spectrum.
Therefore the idling filter also protects the X gate, with $\Delta/2\pi \approx 0.8$~MHz well within the degenerate regime.

For Z gate with $\omega/2\pi=20$~GHz, we have $\mathcal{B}/2\pi \approx 468$~MHz and $\Delta/2 \pi \approx 27$~MHz.
A different filter is required with center frequency at about 20~GHz.
Z gate is unprotected, since the largest possible cooling rate $4\mathcal{B}/25 \approx 75$~MHz is comparable with $\Delta$.

The full protection of the X gate can be verified numerically, and the gate infidelity is reduced by about 2 orders of magnitude with the filter (Fig.~\ref{fig:fig3}(d) orange diamond and dot).
The partial protection of the qubit basis states during the Z gate is verified in Appendix~\ref{SI_gate}.
In principle, the XX gate should also be fully protected for $\alpha_{xx}$ on the order of a few MHz, since its $\Delta$ and $\mathcal{B}$ are similar to the X gate. However, we didn't simulate the XX gate due to the high computational cost.

\begin{figure}[t]
    \centering
    \includegraphics[width=0.48\textwidth]{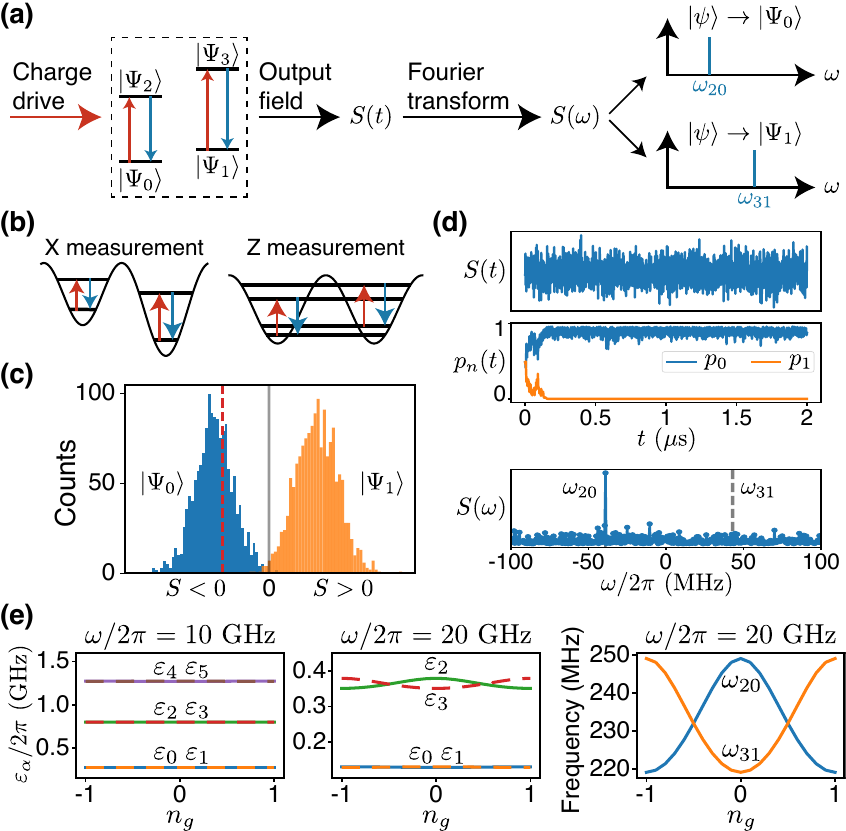}
    \caption{(a) Kapitzonium measurement by measuring the frequency of the emitted photon. (b) The measurement basis depends on how to lift the degeneracy between $\omega_{20}$ and $\omega_{31}$. (c) Quantum trajectory simulation of the Kapitzonium measurement. (d) A single trajectory corresponding to the red dashed line in (c), where $S(\omega)$ is plotted in the rotating frame of $\omega_f$. (e) Charge sensitivity of the Floquet eigenenergies and the emission spectrum.}
    \label{fig:fig4}
\end{figure}

\subsection{Fluorescence-based state measurement}
Kapitzonium measurements can be performed in the unprotected regime. By measuring the frequency of the emitted photon, we learn about which transition it comes from and randomly project the system to either $\ket{\Psi_0}$ or $\ket{\Psi_1}$ (Fig.~\ref{fig:fig4}(a)).
We could apply the X gate but with a much larger $\alpha_x$ to implement the X measurement (Fig.~\ref{fig:fig4}(b)). Larger $\alpha_x$ lifts the degeneracy between $\omega_{20}$ and $\omega_{31}$, and $\{\ket{0},\ket{\pi}\}$ are the measurement basis.
On the other hand, the unprotected Z gate dephases the qubit and naturally perform the Z measurement with measurement basis $\{ \ket{\Psi_0},\ket{\Psi_1} \}$ (Fig.~\ref{fig:fig4}(b)).

The measurement rate is proportional to the occupation of the excited states $\ket{\Psi_2}$ and $\ket{\Psi_3}$.
We could increase the measurement rate by capacitively driving $\ket{\Psi_0} \leftrightarrow \ket{\Psi_2}$ and $\ket{\Psi_1} \leftrightarrow \ket{\Psi_3}$ at the heating transition frequency (Fig.~\ref{fig:fig4}(a)).
The charge operator in the Floquet frame is
\begin{equation}\label{eq:charge_op}
    \begin{split}
        \opn(t) =& \ket{\Psi_0} \bra{\Psi_2} \left( n_{02} e^{i\omega_{02} t} + n_{20} e^{-i \omega_{20} t} + ... \right) \\
        & + \ket{\Psi_1} \bra{\Psi_3} \left( n_{13} e^{i\omega_{13} t} + n_{31} e^{-i \omega_{31} t} + ... \right) + \text{h.c.},
    \end{split}
\end{equation}
where we only include $\ket{\Psi_{0\sim 3}}$ terms from Eq.~(\ref{eq:op_expansion}) and the dominant Fourier coefficients $n_{02} \approx n_{20} \approx n_{13}  \approx n_{31}  \approx 1$.
We apply a charge drive $2\Omega (\cos(\omega_{d1}t) + \cos(\omega_{d2}t)) \opn(t)$ to the Kapitzonium where $\omega_{d1}=\omega_{02}$, $\omega_{d2}=\omega_{13}$ and $\Omega \ll \Delta, \kappa_c$.
This leads to the coupling $\Omega (n_{02} \ket{\Psi_0} \bra{\Psi_2} + n_{13} \ket{\Psi_1} \bra{\Psi_3} + \text{h.c.})$ after the rotating wave approximation (RWA).
Furthermore, $\{ \ket{\Psi_2}, \ket{\Psi_3} \}$ are only weakly excited since $\Omega \ll \kappa_c$, which decay back to $\{ \ket{\Psi_0}, \ket{\Psi_1} \}$ after the measurement with the charge drive turned off.

We simulate the Kapitzonium measurements with quantum trajectory methods~\cite{wiseman2009} (see Appendix~\ref{SI_measurement}).
Starting from an initial state of $(\ket{\Psi_0}+\ket{\Psi_1})/\sqrt{2}$, we monitor the output field from the filter with a heterodyne measurement for $2~\mu \text{s}$, and then calculate the power spectrum $S(\omega)$ for each measurement record $S(t)$.
Depending on whether the system is projected into $\ket{\Psi_0}$ or $\ket{\Psi_1}$, $S(\omega)$ show a peak at $\omega_{20}$ or $\omega_{31}$ (Fig.~\ref{fig:fig4}(a)).
We therefore define the signal as $S=S_{31}-S_{20}$ where $S_{ij}$ is the integrated power within a narrow frequency window around $\omega_{ij}$. $S<0$ or $S>0$ represents a measurement result of 0 or 1.
The measurement fidelity is about 99.4\%, estimated from a total of 3000 trajectories (Fig.~\ref{fig:fig4}(c)).
We plot a single trajectory in Fig.~\ref{fig:fig4}(d) showing the measurement result $S(t)$, the occupation $p_i$ of $\ket{\Psi_i}$ during the measurement for $i=0,1$, and the spectrum $S(\omega)$.

\section{Towards experimental implementation}
\label{main_section_4}

\subsection{Parameter disorder}
Ideally, the SQUID in Fig.~\ref{fig:fig1}(b) should be symmetric to engineer the Kapitzonium Hamiltonian Eq.~(\ref{eq:Hamiltonian_general}).
However in actual experiments, there always exists some amount of disorder which requires a more general treatment of the circuit.

The circuit Hamiltonian in presence of parameter disorder is~\cite{you2019,riwar2022}
\begin{equation}
    \begin{split}
        \oph (t) = 4 E_C \opn^2 & - E_{J1} \cos \left(\opphi - \frac{C_2}{C_{\Sigma}} \Phi_{\text{ext}}(t) \right) \\
        & - E_{J2} \cos \left(\opphi + \frac{C_1}{C_{\Sigma}} \Phi_{\text{ext}}(t) \right) ,
    \end{split}
\end{equation}
where $C_{\Sigma} = C_1+C_2$ is the total junction capacitance and $E_C = e^2/2C_{\Sigma}$.
To symmetrize the flux allocation inside the two cosines, we move to another reference frame by performing the unitary transformation
\begin{equation}
    \opu = \exp \left( i \opn \frac{C_2-C_1}{2 C_\Sigma} \Phi_{\text{ext}}(t) \right) ,
\end{equation}
where the Hamiltonian becomes
\begin{equation}
    \begin{split}
        \oph (t) =& 4 E_C \opn^2 -  E_{J1} \cos \left(\opphi - \frac{1}{2} \Phi_{\text{ext}}(t) \right) \\
        &-  E_{J2} \cos \left(\opphi + \frac{1}{2} \Phi_{\text{ext}}(t) \right) - \frac{C_2-C_1}{2 C_\Sigma} \dot{\Phi}_{\text{ext}}(t) \opn .
    \end{split}
\end{equation}
We will drop the unwanted term $\propto \dot{\Phi}_{\text{ext}}(t) \opn$ for now, since in principle it can be compensated for with the gate voltage.

For disorder in junction energies, we define $\delta_e = (E_{J1}-E_{J2})/(E_{J1}+E_{J2})$.
With the idling flux drive $\phi_{\text{ext}}(t) = \omega t$, the effective Hamiltonian is
\begin{equation}
    \oph_{\text{eff}} = 4 E_C \opn^2 - \tilde{E}_J (1-\delta_e^2) \cos 2\opphi .
\end{equation}
Therefore disorder in junction energies reduces the effective $\tilde{E}_J$, and for small $\delta_e$ on the order of 0.05 this reduction is likely negligible.

\subsection{Flux control}
\label{sec:flux_drive}
Here we discuss the flux control for implementing the Kapitzonium gates and measurements.
The flux drives $\phi_{\text{ext}}(t)=\omega t$ (idling) and $\phi_{\text{ext}}(t)=\omega_z t$ (Z gate and Z measurement) are not feasible experimentally, since the required bias current grows linearly with time.
One natural solution is to use a triangle waveform instead
\begin{equation}
    \tilde{\phi}_{\text{ext}}(t) = \left \{
    \begin{array}{ll}
        u(t) & 0 \leq u(t) < 2\pi \\
        4\pi - u(t) & 2\pi \leq u(t) < 4\pi
    \end{array}
    \right . ,
\end{equation}
where $u(t) = \phi_{\text{ext}}(t) \pmod{4\pi}$. However, this flux choice poses inconveniences for implementing the X gate and X measurement.

Alternatively, we could set $\phi_{\text{ext}}(t)=\alpha \cos \omega t$ with
\begin{equation}
    \cos \phi_{\text{ext}}(t) = J_0(\alpha) + 2\sum_{n=1}^\infty (-1)^n J_{2n}(\alpha) \cos 2n\omega t ,
\end{equation}
where we have applied the Jacobi–Anger expansion and $J_n(x)$ is the $n$-th Bessel function of the first kind.
The effective Hamiltonian for Eq.~(\ref{eq:Hamiltonian_general}) now becomes
\begin{equation}
    \begin{split}
        \oph_{\text{eff}} =& 4 E_C \opn^2 - E_J J_0(\alpha) \cos \opphi \\
        & - \tilde{E}_J \sum_{n=1}^\infty \left( \frac{J_{2n} (\alpha)}{n} \right)^2 \cos 2\opphi .
    \end{split}
\end{equation}
Therefore we could choose different $\alpha$ such that $J_0(\alpha)=0$ for Kapitzonium idling, Z gate and Z measurement, and $J_0(\alpha) \neq 0$ for X gate and X measurement.

\subsection{Coherence properties}
We first estimate the coherence properties of the Kapitzonium, while neglecting any nonidealities due to flux, quasiparticle and offset charge noise.
In this idealized model, both bit-flip rate $1/T_1$ and phase-flip rate $1/T_2$ of the Kapitzonium without cooling are 0, since at any time $t$
\begin{equation}
    \mele{\Psi_\alpha (t)}{\opn}{\Psi_\beta (t)} = 0
\end{equation}
for $\alpha,\beta \in \{0,1\}$.
This is because the Kapitzonium Hamiltonian (Eq.~(\ref{eq:Hamiltonian_idle})) is symmetric under the transformation of $\opn \rightarrow -\opn$, and both $\ket{\Psi_0(t)}$ and $\ket{\Psi_1(t)}$ have even parity under this transformation.
More generally with a static offset charge, numerical results suggest that both bit-flip and phase-flip rates are exponentially small in $E_J/\omega$ (see Appendix~\ref{SI_lifetime_ng}).
Due to the Floquet driving, the dominant error is the leakage from the qubit subspace, e.g., from $\ket{0}(\ket{\pi})$ to $\ket{\psi_2} (\ket{\psi_3})$. This leakage rate $1/T_l$ is proportional to $\kappa_h$.

The filter implemented in Sec.~\ref{main_section_3} suppresses the leakage rate by $\kappa_h/\kappa_c$ so $1/T_l\propto \kappa_h^2/\kappa_c$.
Such a cooling process induces dephasing error due to the frequency difference $\Delta$ in the emitted photons, with a rate of $1/T_2 \propto \frac{\kappa_h}{2} \left( \frac{\Delta}{\kappa_c} \right)^2$. In the deep transmon regime $E_J \gg \omega$, we have $\Delta \propto \exp(-\frac{\pi^2 E_J}{8 \sqrt{2} \omega})$ and the dephasing rate is exponentially small in $E_J/\omega$.

In summary, the bit and phase flip rates are exponentially suppressed, while the leakage rate is reduced by the filter factor $\kappa_h/\kappa_c$. Below, we give a preliminary discussion of how other noise processes that affect superconducting circuits further reduce the coherence of the Kapitzonium.

\textbf{Flux noise}.
The flux imposed on the loop will be invariably accompanied by additional flux noise $\delta \phi$, which typically has a $f^{-\alpha}$ spectral density and thus a small spectral weight at high frequencies.
The effects of $\delta \phi$ depend on how we choose to drive the flux.
For $\phi_{\text{ext}}(t)=\omega t + \delta \phi \pmod{4\pi}$, $\delta \phi$ corresponds to a shift in time which does not change the Floquet eigenenergies.
Since the idling, Z gate and Z measurement can be implemented in this way (Sec.~\ref{sec:flux_drive}), they are robust to flux noise.
However, experimental constraints in implementing a flux drive, as well as the proposed approach to X gate and X measurement, make other driving schemes such as $\phi_{\text{ext}}(t)=\alpha \cos \omega t$ (Sec.~\ref{sec:flux_drive}) more attractive. In this case, flux noise leads to an imposed flux of $\phi_{\text{ext}}(t)=\alpha \cos \omega t + \delta \phi$ where $\delta \phi$ has a nontrivial effect on the effective Hamiltonian.

Flux sensitivity at arbitrary flux bias occurs in other protected qubits subject to experimental constraints. For example it is also present in the static soft $0-\pi$ qubit~\cite{groszkowski2018,gyenis2021}. Nonetheless, as in the case of the soft  $0-\pi$ qubit~\cite{gyenis2021}, it is possible to operate at flux sweet-spot where the dephasing rate is protected to first order against flux noise.
For a Kapitzonium with a noisy periodic flux drive $\phi_{\text{ext}}(t)=\alpha \cos \omega t + \delta \phi$, the effective Hamiltonian is also insensitive to first order to $\delta \phi$ around $\delta \phi=0$, and leading order terms scale as $\delta \phi^2$. Therefore operating at the sweet spot  $\delta \phi=0$ can significantly reduce errors of X gate and X measurement.

\textbf{Quasiparticle noise}.
Quasiparticle tunneling changes the charge parity. This may induces both energy relaxation and dephasing~\cite{lutchyn2005,lutchyn2006,koch2007,martinis2009,catelani2011, serniak2018} of the Kapitzonium. Here we note that in a highly simplified model of quasiparticle dynamics, the frequency shifts in the $\{ \ket{0},\ket{\pi} \}$ basis induced by quasiparticles are proportional to $\mele{0}{\sin \frac{\hat{\varphi}_k}{2}}{0}\approx (-1)^{k-1} \sin \frac{\phi_{\text{ext}}(t)}{2}$, where $k=1,2$ is the index of the junction, and $\mele{\pi}{\sin \frac{\hat{\varphi}_k}{2}}{\pi}\approx \cos \frac{\phi_{\text{ext}}(t)}{2}$. Therefore for $\phi_{\text{ext}}(t)=\omega t$, $\ket{0}$ and $\ket{\pi}$ on average don't accumulate extra phase and should be robust to quasiparticle tunneling
. This analysis may not hold when $\phi_{\text{ext}}(t)=\alpha \cos \omega t$. A more accurate model of quasiparticle tunneling is also likely to be needed to better understand the effects of quasiparticles in the Kapitzonium. In particular, the modulation drive itself can cause photon-assisted effects~\cite{houzet2019}. We leave such a detailed study to future work.

\textbf{Offset charge noise}.
In addition to quasiparticle tunneling which flips the parity of gate charge, the Kapitzonium, like the charge qubit from which it is derived, suffers from the continuous variations in the gate charge offset parameter due to uncontrolled fluctuations in the background electric fields.
This can be modeled by replacing $\opn$ with $\opn-n_g$ in the Kapitzonium Hamiltonian where $n_g$ is the random offset charge.

The Floquet eigenenergies weakly depend on $n_g$ during idling (Fig.~\ref{fig:fig4}(e), left figure).
The weak dependence also holds when adding $\alpha_x \cos \opphi$ to the Hamiltonian.
Therefore idling, X gate and X measurement are all insensitive to charge noise.
On the other hand, Z gate and Z measurement are not robust to charge noise since $\varepsilon_1 - \varepsilon_0$ and the emitted photon frequency $\omega_{20},\omega_{31}$ depend on $n_g$ (Fig.~\ref{fig:fig4}(e), middle and right figures).
For example, at $n_g=0.5$ we have $\varepsilon_1 = \varepsilon_0$ and $\omega_{20}=\omega_{31}$. Therefore both Z gate and Z measurement cannot be performed at $n_g=0.5$.

Finding ways to work around these challenges as well as other issues which we may learn about from experiments (such as photon-assisted tunneling and unforeseen effects due to the drive), will be the subject of future work.

\section{Discussion}
\label{main_section_5}
We have proposed a quantum Kapitza pendulum in superconducting circuit as a Floquet $0-\pi$ qubit. We identify how single- and two-qubit gates can be implemented, and propose a cooling scheme to protect the Kapitzonium against charge noise. Remarkably, we find that this exceedingly simple Floquet superconducting circuit, a flux-modulated capacitively-shunted SQUID loop, can support a protected qubit subspace.
Our work reveals some of the subtle features of Floquet qubits -- we elucidate the challenges associated with noise-induced heating, and how they can be overcome using filter cavities, and even used to our advantage to realize a fluorescence-based method for qubit state measurement.
Our work lays the groundwork to study new Floquet systems for quantum information processing with superconducting circuits, and outlines a path towards experiments with such devices.

\begin{acknowledgments}
We thank Yudan Guo, Taha Rajabzadeh, Nathan Lee, Takuma Makihara, Qile Su, Jayameenakshi Venkatraman and Jeremy Boaz Kline for helpful discussions. This work was supported by 
the U.S. government through the Office of Naval Research (ONR) under grant No. N00014-20-1-2422 and the National Science Foundation CAREER award No.~ECCS-1941826, and by Amazon Web Services Inc. A.H.S.-N. acknowledges support from the Sloan fellowship. 
\end{acknowledgments}

\appendix

\section{Floquet master equation}
\label{SI_master_eq}
In this section, we derive the Floquet master equation~\cite{grifoni1998} and the Kapitzonium dissipator.
The system-bath Hamiltonian $\oph(t) = \hat{O}(t) \otimes \hat{B}(t)$ in the interaction picture generates the dynamics
\begin{equation}
    \begin{split}
        & \ddt \hat{\rho}_{SB}(t) = -i [\hat{H}(t), \hat{\rho}_{SB}(t)] \\
        =& -i \left[ \hat{H}(t), \hat{\rho}_{SB}(0) - i \int_{0}^t [\hat{H}(\tau), \hat{\rho}_{SB}(\tau)] \text{d}\tau \right] \\
        =& -i [ \hat{H}(t), \hat{\rho}_{SB}(0) ] - \left[ \hat{H}(t), \int_{0}^t [\hat{H}(\tau), \hat{\rho}_{SB}(\tau)] \text{d}\tau \right] .
    \end{split}
\end{equation}
Now we make the Born approximation $\hat{\rho}_{SB}(t) = \hat{\rho}(t) \otimes \hat{\rho}_B$, where $\hat{\rho}(t)$ is the system density matrix and $\hat{\rho}_B$ is the stationary bath density matrix.
In addition, we also make the standard assumption that $\tr [\hat{B}(t) \hat{\rho}_B] = 0$.
The system dynamics becomes
\begin{equation}\label{eq:before_markov}
    \begin{split}
        & \ddt \hat{\rho}(t) = - \tr_B \left[ \hat{H}(t), \int_{0}^t [\hat{H}(t-\tau), \hat{\rho}(t-\tau) \otimes \hat{\rho}_B] \text{d}\tau \right] \\
        =& \int_{0}^t \left( \hat{O} (t-\tau) \hat{\rho} (t-\tau) \hat{O}(t) - \hat{O}(t) \hat{O} (t-\tau) \hat{\rho} (t-\tau) \right) \\
        & \qquad C_B(t,t-\tau) \text{d}\tau  + \text{h.c.} ,
    \end{split}
\end{equation}
where we define the two-point correlation functions
\begin{equation}
    \begin{split}
        & C_B(t,t-\tau) = \ave{\hat{B} (t) \hat{B} (t-\tau)} \\
        =& \tr_B [\hat{B} (t) \hat{B} (t-\tau) \hat{\rho}_B] = C_B(t-\tau,t)^* .
    \end{split}
\end{equation}
To proceed, we assume that the bath is stationary with a very short correlation decay time. In other words
\begin{equation}
    C_B(t,t-\tau) = C_B(\tau,0) \equiv C_B(\tau) \sim e^{-\tau/\tau_B}
\end{equation}
where $\tau_B$ is much shorter than any time scale we are interested in.
Assuming weak system-bath coupling, we have $|\hat{O}(t) - \hat{O}(t-\tau_B)|\sim |\hat{O}|$ and $|\hat{\rho}(t) - \hat{\rho}(t-\tau_B)|\sim |\hat{O}|^2$ which is second order in the coupling strength. 
Therefore we could make the Markov approximation and replace $\hat{\rho}(t-\tau)$ with $\hat{\rho}(t)$ in Eq.~(\ref{eq:before_markov}).
Since only $\tau \approx 0$ contribute significantly to the integration, the upper limit of the integration can be extended to $\infty$:
\begin{equation}\label{eq:Born-Markov}
    \begin{split}
        \ddt \hat{\rho}(t) =& \int_{0}^\infty \left( \hat{O} (t-\tau) \hat{\rho} (t) \hat{O}(t) - \hat{O}(t) \hat{O} (t-\tau) \hat{\rho} (t) \right) \\
        & \qquad C_B(\tau) \text{d}\tau  + \text{h.c.} .
    \end{split}
\end{equation}

For Floquet systems, we have $\hat{O}(t) = \sum_{\omega} \hat{O}(\omega) e^{-i\omega t}$ with $\hat{O}(\omega) = \hat{O}^\dagger (-\omega)$ since $\hat{O}(t)$ is Hermitian. Therefore
\begin{equation}
    \int_{0}^\infty \hat{O} (t-\tau) C_B(\tau) \text{d}\tau = \sum_{\omega} \hat{O}(\omega) e^{-i\omega t} \Gamma (\omega) ,
\end{equation}
where
\begin{equation}
    \Gamma (\omega) = \int_{0}^\infty e^{i\omega \tau} C_B(\tau) \text{d}\tau .
\end{equation}
We could decompose $\Gamma(\omega)$ into its real and imaginary parts as $\Gamma(\omega) = \frac{1}{2} \gamma (\omega) + i S(\omega)$, where
\begin{equation}
    \begin{split}
        \gamma (\omega) =& \int_{-\infty}^\infty e^{i\omega \tau} C_B(\tau) \text{d}\tau \\
        S (\omega) =& \int_{-\infty}^\infty \frac{\text{d} \omega'}{2\pi} \gamma(\omega') \mathcal{P} \left( \frac{1}{\omega - \omega'} \right) .
    \end{split}
\end{equation}
Physically the real part $\gamma(\omega)$ represents the decay rate while the imaginary part $S (\omega)$ can be absorbed into the system Hamiltonian which we will ignore for now.

Consider a zero temperature bath with a flat spectral density function
\begin{equation}\label{eq:flat_spec}
    \gamma(\omega) = \left \{
    \begin{array}{ll}
        \gamma & \quad \omega > 0 \\
         0 & \quad \omega \leq 0
    \end{array} \right.
\end{equation}
as an example.
We could decompose $\hat{O}(t)$ into positive and negative frequency parts where
\begin{equation}
    \hat{O}_+ (t) = \sum_{\omega<0} \hat{O}(\omega) e^{-i\omega t} \qquad \hat{O}_- (t) = \sum_{\omega>0} \hat{O}(\omega) e^{-i\omega t} .
\end{equation}
and $\hat{O}_+ (t) = \hat{O}^\dagger_{-} (t)$.
Therefore 
\begin{equation}
    \int_{0}^\infty \hat{O} (t-\tau) C_B(\tau) \text{d}\tau = \frac{\gamma}{2} \sum_{\omega>0} \hat{O}(\omega) e^{-i\omega t} = \frac{\gamma}{2} \hat{O}_- (t) ,
\end{equation}
and Eq.~(\ref{eq:Born-Markov}) gives the Floquet master equation
\begin{equation}
    \begin{split}
        \ddt \hat{\rho}(t) =& \frac{\gamma}{2} \hat{O}_- (t) \hat{\rho} (t) \hat{O}(t) - \frac{\gamma}{2} \hat{O}(t) \hat{O}_- (t) \hat{\rho} (t)  + \text{h.c.} \\
         =& \gamma D[\hat{O}_- (t)] \hat{\rho} (t) ,
    \end{split}
\end{equation}
where we apply RWA to drop $\hat{O}^2_{\pm} (t)$ terms.
This justifies the Kapitzonium dissipator Eq.~(\ref{eq:neg_frequency}) in the main text.

\section{Gate simulation}
\label{SI_gate}
\subsection{Average gate fidelity}
We benchmark the Kapitzonium gates with the measure of average fidelity $\bar{F}$~\cite{nielsen2002,wang2022c}.
Intuitively, $\bar{F}$ describes how well the qubit subspace is preserved under some quantum process.
More specifically, the average fidelity of a quantum channel $\mathcal{M}$ over all states $\ket{\psi} = \sum_{n=1}^N c_n \ket{n}$ in a $N$-dimensional qubit subspace is
\begin{equation}
    \begin{split}
        \bar{F} =& \int d\psi \tr \left[ \mathcal{M} (\ket{\psi} \bra{\psi}) \ket{\psi} \bra{\psi} \right] \\
        =& \sum_{mn} \tr \left[ \mathcal{M} (\ket{m} \bra{n}) \rho_{mn} \right] ,
    \end{split}
\end{equation}
where
\begin{equation}
    \begin{split}
        \rho_{mn} =& \int d\psi c_m c_n^* \ket{\psi} \bra{\psi} = \sum_{kl} \ket{k} \bra{l} \int d\psi c_m c_n^* c_k c_l^* \\
        =& \delta_{mn} \left[ \frac{2}{N(N+1)} \ket{n} \bra{n} + \sum_{k \neq n} \frac{1}{N(N+1)} \ket{k} \bra{k} \right] \\
        &+ (1-\delta_{mn}) \frac{1}{N(N+1)} \ket{n} \bra{m} .
    \end{split}
\end{equation}
Therefore the average fidelity can be simplified as
\begin{equation}
    \begin{split}
        \bar{F} =& \frac{1}{N(N+1)} \sum_{n} \tr \left[ \mathcal{M} (\ket{n} \bra{n}) \sum_{k} (1+\delta_{nk}) \ket{k} \bra{k} \right] \\
        &+ \frac{2}{N(N+1)} \sum_{m<n} \text{Re} \left\{ \tr \left[ \mathcal{M} (\ket{m} \bra{n}) \ket{n} \bra{m} \right] \right\} .
    \end{split}
\end{equation}
To compare $\mathcal{M}$ with some target unitary $\hat{U}$, we could compare $\mathcal{M}'$ with identity instead, where $\mathcal{M}'(\hat{\rho}) \equiv \hat{U}^\dagger \mathcal{M} (\hat{\rho}) \hat{U}$.

\subsection{Unitary case}
The gates are designed such that the system Hamiltonian adiabatically evolve from the idling $\oph_0(t)$ to the gate Hamiltonian $\oph_{\text{gate}}(t)$, and then adiabatically evolve back to $\oph_0(t)$ after certain amount of gate time.
The average gate fidelity is calculated for the single-qubit subspace with $N=2$ or the two-qubit subspace with $N=4$.

We choose the pulse shape
\begin{equation}
    \alpha (t,t_{\text{gate}},\tau) = \left \{
    \begin{array}{ll}
        \sin \left( \frac{\pi t}{2\tau} \right)^2 & 0 \leq t < \tau \\
        1 & \tau \leq t \leq t_{\text{gate}} - \tau \\
        \sin \left( \frac{\pi(t_{\text{gate}}-t)}{2\tau} \right)^2 & t_{\text{gate}} - \tau < t \leq t_{\text{gate}} 
    \end{array}
    \right . ,
\end{equation}
where $t_{\text{gate}}$ is the total gate duration and $\tau$ is the adiabatic ramping time.

\textbf{X gate}.
The Hamiltonian is
\begin{equation}
    \hat{H}_x(t) = \oph_0(t) + \alpha_x \alpha(t,t_x,\tau_x) \cos \hat{\phi} ,
\end{equation}
where $t_x=60~$ns, $\tau_x=10~$ns and $\alpha_x/2\pi\approx 5.2~$MHz.
The X gate implements the mapping of $\ket{\Psi_0} \rightarrow \ket{\Psi_1},\ket{\Psi_1} \rightarrow \ket{\Psi_0}$ with an infidelity of $1.7 \times 10^{-7}$.

\textbf{Z gate}.
The Hamiltonian is
\begin{equation}
    \hat{H}_z(t) = 4 E_C \hat{n}^2 - E_{J} \alpha_z (t) \cos \hat{\phi} ,
\end{equation}
where
\begin{equation}
    \alpha_z (t) = (1-\alpha(t,t_z,\tau_z)) \cos \omega t + \alpha(t,t_z,\tau_z) \cos \omega_z t .
\end{equation}
We choose $t_z=296.2~$ns, $\tau_z=20~$ns and $\omega_{z}/2\pi=20$~GHz. The Z gate implements the mapping of $\ket{\Psi_0} \rightarrow \ket{\Psi_0},\ket{\Psi_1} \rightarrow -\ket{\Psi_1}$ with an infidelity of $4.6 \times 10^{-6}$.

Another Floquet drive that seems reasonable at first is to have frequency modulation instead of amplitude modulation:
\begin{equation}
    \alpha_z (t) = \cos [ ((1-\alpha(t,t_z,\tau_z)) \omega + \alpha(t,t_z,\tau_z) \omega_z)t ] .
\end{equation}
However, this Floquet drive always leads to unstable dynamics which heats up the Kapitzonium even with very slow ramping.

\textbf{XX gate}.
The Hamiltonian is
\begin{equation}
    \hat{H}_{xx}(t) = \hat{H}_{0}^{(1)}(t) + \hat{H}_{0}^{(2)}(t) + \alpha_{xx} \alpha(t,t_{xx},\tau_{xx}) \cos (\hat{\phi}_1 - \hat{\phi}_2) ,
\end{equation}
where $\hat{H}_0^{(i)}(t) = 4 E_{C} \hat{n}_i^2 - E_{J} \cos \omega t \cos \hat{\phi}_i$ is the idling Hamiltonian for each qubit with $i=1,2$.
We choose $t_{xx}=39~$ns, $\tau_{xx}=12~$ns and $\alpha_{xx}/2\pi=10$~MHz. The XX gate implements the mapping of $\ket{\Psi_0 \Psi_0} \rightarrow \ket{\Psi_1 \Psi_1}, \ket{\Psi_0 \Psi_1} \rightarrow \ket{\Psi_1 \Psi_0}, \ket{\Psi_1 \Psi_0} \rightarrow \ket{\Psi_0 \Psi_1}, \ket{\Psi_1 \Psi_1} \rightarrow \ket{\Psi_0 \Psi_0}$ with an infidelity of $2.4 \times 10^{-7}$.

\textbf{State initialization}.
To initialize the system state into $\ket{0}$, we could start from the ground state of the static transmon Hamiltonian $\oph = 4 E_C \hat{n}^2 - E_{J} \cos \hat{\phi}$, and adiabatically apply the Floquet drive:
\begin{equation}
    \hat{H}(t) = 4 E_C \hat{n}^2 - E_{J} \left( \alpha(t) \cos \omega t + (1-\alpha(t)) \right) \cos \hat{\phi} ,
\end{equation}
where $\alpha(t)$ increases from 0 to 1.

\subsection{Open system without filter}
In the open system simulation without filter, the Floquet drives are the identical to the unitary case.
The only difference is that during $t\in [\tau, t_{\text{gate}}-\tau]$ we add loss to the system. More specifically, the simulation is performed in the Floquet frame with Hamiltonian 0 and a single Lindblad dissipator $\kappa_h D[\hat{O}_- (t)]$. Notice that the Floquet frame here is defined with respect to $\oph_{\text{gate}}(t)$ instead $\oph_0(t)$.

During the ramping parts of Floquet drive, the Hamiltonian is not strictly time-periodic which makes it difficult to calculate the time-dependent dissipator.
Therefore the ramping parts are always assumed to be unitary where the simulation is done in the lab frame.

\subsection{Open system with filter}
Due to the hybridization between the Kapitzonium and the filter, $\ket{\Psi_\alpha (t)} \otimes \ket{0_f}$ is no longer the Floquet eigenstates of full Hamiltonian (Eq.~(\ref{eq:H_with_filter}) in the main text) where $\ket{0_f}$ is the ground state of the filter.
Therefore we work with the dressed Floquet eigenstates $\widetilde{\ket{\Psi_\alpha (t)}}$ of Eq.~(\ref{eq:H_with_filter}) instead.
The qubit basis states $\{ \widetilde{\ket{\Psi_0 (t)}}, \widetilde{\ket{\Psi_1 (t)}} \}$ are chosen based on their overlap with $\{ \ket{\Psi_0 (t)} \otimes \ket{0_f}, \ket{\Psi_1 (t)} \otimes \ket{0_f} \}$.

The Floquet drive parameters requires a slight fine tuning due to this hybridization.
During the gate time $t\in [\tau, t_{\text{gate}}-\tau]$, the simulation is performed in the Floquet frame defined by $\oph_{\text{gate}}(t)$ with Hamiltonian 0 and two Lindblad dissipators $\kappa_h D[\hat{O}_- (t)]$ and $\kappa_f D[\hat{a}_N(t)]$. Here $\hat{a}_N(t)$ is calculated similarly to Eq.~(\ref{eq:neg_frequency}) with $\hat{O} = \hat{a}_N + \hat{a}_N^\dagger$.
To remove any transient effects at the beginning of the cooling~\cite{kapit2018}, we prepare the initial states for benchmarking the gates by evolving the qubit basis states for 50~ns idling until the system reaches equilibrium.

\begin{figure}[t]
    \centering
    \includegraphics[width=0.48\textwidth]{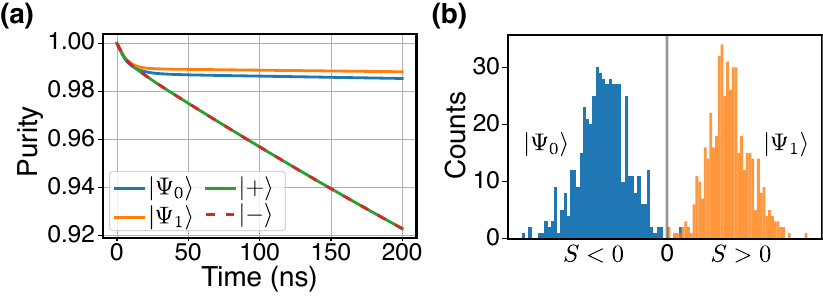}
    \caption{(a) Z gate simulation with filter for different initial states. (b) Z measurement results.}
    \label{fig:SI_fig_1}
\end{figure}

In Fig.~\ref{fig:SI_fig_1}(a), we simulate the unprotected Z gate for different initial states.
Here $\ket{\Psi_0}$ and $\ket{\Psi_1}$ are the dressed Floquet eigenstates with $\omega/2\pi=20$~GHz and $\ket{\pm}$ are their even and odd superpositions.
We choose $\omega_{f}/2\pi \approx 20.234~$GHz, $\kappa_f/2\pi=200~$MHz $J=\kappa_f/2$ and $g=\kappa_f/5$ for the Z gate filter.

We use QuTiP~\cite{johansson2012,johansson2013} for all the simulations, and modify the built-in \texttt{mesolve} function to speed up the open system simulation with time dependent dissipators.

\section{Measurement simulation}
\label{SI_measurement}
The Hamiltonian for the measurement simulation is
\begin{equation}
    \begin{split}
        \oph (t) =& 2\Omega (\cos(\omega_{d1}t) + \cos(\omega_{d2}t)) \opn(t) \\
        & + g \opn(t) (\hat{a}_1 e^{-i\omega_{f} t} + \hat{a}_1^\dagger e^{i\omega_{f} t}) \\
        & + J \sum_{k=1}^{N-1} \left( \hat{a}_k \hat{a}_{k+1}^\dagger + \hat{a}_k^\dagger \hat{a}_{k+1} \right) ,
    \end{split}
\end{equation}
which is in the Floquet frame of the Kapitzonium and the rotating frame of the filter modes.
Notice that the simulation only includes the first 4 Floquet eigenstates with $\opn(t)$ in Eq.~(\ref{eq:charge_op}) represented by a $4\times 4$ matrix.

For X measurement, the system Hamiltonian is $\oph_x = \oph_0(t) + \alpha_x \cos \opphi$ with $\alpha_x/2\pi=500$~MHz.
$\opn(t)$ can be calculated from the emission spectrum of $\oph_x$. The charge drive frequencies are $\omega_{02}/2\pi \approx 9.44~$GHz and $\omega_{13}/2\pi \approx 9.52~$GHz. The emitted photon frequencies are $\omega_{20}/2\pi \approx 10.56~$GHz and $\omega_{31}/2\pi \approx 10.48~$GHz.
We choose $\Omega/2\pi=6.4~$MHz and the filter parameters $g,J,\omega_{f}$ are the same as the idling filter.

For Z measurement, the system Hamiltonian is the same as the Z gate with an additional charge drive.
The charge drive frequencies are $\omega_{02}/2\pi \approx 19.78~$GHz and $\omega_{13}/2\pi \approx 19.75~$GHz. The emitted photon frequencies are $\omega_{20}/2\pi \approx 20.22~$GHz and $\omega_{31}/2\pi \approx 20.25~$GHz.
We choose $\Omega/2\pi=2.3~$MHz and the filter parameters are the same as the Z gate filter.
Starting from the initial state $(\ket{\Psi_0}+\ket{\Psi_1})/\sqrt{2}$, we simulate 1000 trajectories for the Z measurement with a measurement time of $10~\mu$s (Fig.~\ref{fig:SI_fig_1}(b)) and the measurement fidelity is about 99.8\%.

We would like to make a few comments on Kapitzonium measurement.
First of all, the charge drive could be slightly off-resonant from the heating transitions with $\omega_{d1} \approx \omega_{02}$, $\omega_{d2} \approx \omega_{13}$, which shifts the emitted photon frequencies as well.
Second of all, the charge drive frequency should be outside the filter passband. In principle setting $\omega_{d1} \approx \omega_{20}$, $\omega_{d2} \approx \omega_{13}$ also drives the Rabi oscillation via the cooling transitions. However, this could cause measurement error if there is any direct leakage from the charge drive to the output of the filter.
Finally, choosing $\Omega$ to be comparable or larger than $\Delta$ will cause measurement error due to crosstalk between the drives. On the other hand, a very small $\Omega$ reduces the measurement rate and requires a long measurement time.

\begin{figure}[t]
    \centering
    \includegraphics[width=0.45\textwidth]{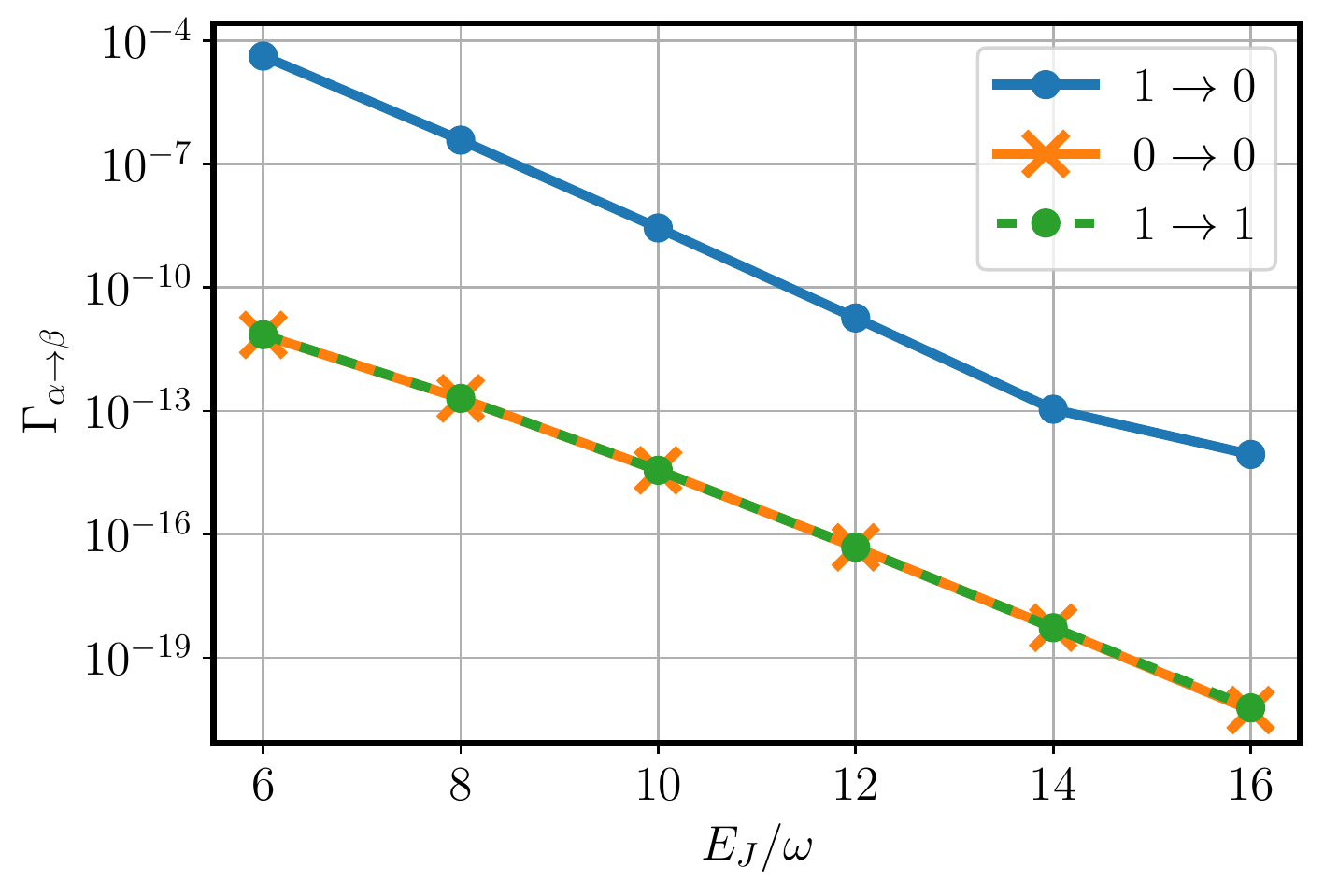}
    \caption{Kapitzonium transition rates for $n_g=0.3$. The junction energies are $E_J/2\pi= 60,80,100,120,140,160$~GHz, with $\omega/2\pi=10~\text{GHz}$ and $E_C/2\pi=0.01~\text{GHz}$ fixed.}
    \label{fig:SI_fig_2}
\end{figure}

\section{Kapitzonium lifetime estimation}
\label{SI_lifetime_ng}
The Kapitzonium Hamiltonian in presence of offset charge $n_g$ is
\begin{equation}
    \oph_0(t) = 4 E_C (\opn-n_g)^2 - E_J \cos \omega t \cos \opphi .
\end{equation}
For $n_g=0$, we have $\hat{\Pi} \oph_0(t) = \oph_0(t) \hat{\Pi}$ where the parity operator is $\hat{\Pi} = \sum_n \ket{-n} \bra{n}$ and $\{ \ket{n} \}$ are the charge eigenstates.
The Floquet eigenstates $\ket{\Psi_\alpha(t)}$ are parity eigenstates, and it turns out that both $\ket{\Psi_0(t)}$ and $\ket{\Psi_1(t)}$ have even parity.
Therefore
\begin{equation}
    \begin{split}
        & \mele{\Psi_\alpha (t)}{\opn}{\Psi_\beta (t)} = \mele{\Psi_\alpha (t)}{\hat{\Pi} \opn \hat{\Pi}}{\Psi_\beta (t)} \\
        =& -\mele{\Psi_\alpha (t)}{\opn}{\Psi_\beta (t)} = 0 ,
    \end{split}
\end{equation}
for $\alpha,\beta \in \{0,1\}$.

For $n_g \neq 0$, $\oph_0(t)$ don't have the symmetry under $\hat{\Pi}$ and the bit-flip and phase-flip rates are no longer exactly 0.
However, since Kapitzonium is in the deep transmon regime, we still expect the differences from the $n_g=0$ results to be exponentially small in $E_J/\omega$.
We define the total transition rate from $\ket{\Psi_\alpha}$ to $\ket{\Psi_\beta}$ as
\begin{equation}
    \Gamma_{\alpha \rightarrow \beta} \equiv \sum_{\varepsilon_\alpha - \varepsilon_\beta +n \omega > 0} |O_{\alpha\beta n}|^2 ,
\end{equation}
where $1/T_1 \sim \Gamma_{1 \rightarrow 0}$ and $1/T_2 \sim \Gamma_{0 \rightarrow 0}, \Gamma_{1 \rightarrow 1}$.
In Fig.~\ref{fig:SI_fig_2}, we plot the transition rates for different values of $E_J/\omega$. The results indeed show an exponential suppression of both the bit-flip and phase flip rates with $E_J/\omega$.

\bibliography{Kapitzonium.bib}

\end{document}